\documentclass[twocolumn,amsmath,amssymb,prl]{revtex4}
\usepackage{graphicx}
\usepackage{epsfig}
\usepackage{amssymb}
\usepackage{amsmath}
\usepackage{color} 
\usepackage{pstricks}
\usepackage[bookmarks=true]{hyperref}

\newcommand{\ket}[1]{\vert{#1}\rangle}
\newcommand{\bra}[1]{\langle{#1}\vert}
\newcommand{\Tr}[1]{\textrm{Tr}\left[{#1}\right]}
\newcommand{\ham}{\mathcal{H}}

\input{./Qcircuit}

\begin{document}
\title {Coherence and control of quantum registers based on  electronic spin in a nuclear spin bath}
\author{P. Cappellaro$^{1,2}$, L. Jiang$^2$, J.~S. Hodges$^{2,3}$ and M.~D. Lukin$^{1,2}$}
\affiliation{$^1$ITAMP -- Harvard-Smithsonian Center for Astrophysics and  $^2$Physics Department, Harvard University, Cambridge, MA 02138, USA. $^3$Department of Nuclear Science and Engineering, Massachusetts Institute of Technology, Cambridge, MA 02139, USA}

\begin{abstract}
We consider a protocol for the control of few-qubit registers comprising one electronic spin embedded in a nuclear spin bath. We show how to isolate a few proximal nuclear spins from the rest of the environment and use them as  building blocks for a potentially scalable quantum information processor.  We describe how coherent control techniques  based on magnetic resonance methods can be adapted to these electronic-nuclear solid state spin systems, to provide not only efficient, high fidelity manipulation of the registers, but also decoupling from the spin bath. As an example, we analyze feasible performances and practical limitations in a realistic setting associated with nitrogen-vacancy centers in diamond. 
\end{abstract}
\maketitle
The coherence properties of  isolated electronic spins  
in solid-state materials are frequently determined by their interactions with large ensembles of lattice nuclear spins \cite{coishLoss,Koppens05Etal}. The dynamics of nuclear spins is typically slow, resulting in very long correlation times of the environment. Indeed, nuclear spins represent one of the most isolated systems available in nature. 
This allows, for instance, to decouple electronic spin qubits from nuclear spins via spin echo techniques \cite{PettaScienceEt,HansonBathSciEt}. Even more remarkably, controlled manipulation of the coupled electron-nuclear system allows one to take advantage of the nuclear spin environment and use  it as a long-lived quantum memory \cite{Taylor03,Mehring06,Morton08}. Recently, this approach has been used to demonstrate a single nuclear qubit memory with coherence times well exceeding tens of milliseconds in room temperature diamond~\cite{Dutt06Et}.  Entangled states composed of one electronic spin and two nuclear spin qubits have been probed in such a system \cite{Neumann08Et}. The essence of these experiments is to gain complete control over several  nuclei by isolating them from the rest of the nuclear spin bath. Universal control of systems comprising a single electronic spin coupled to many nuclear spins has not been demonstrated yet and could enable developing of robust quantum nodes to build larger scale quantum information architectures. 

In this Letter, we describe a technique to achieve  coherent universal control of a portion of the nuclear spin environment. In particular,  we show how several of these nuclear spins 
can be used, together with an electronic spin, to build robust multi-qubit quantum registers.  
Our approach is based on quantum control techniques associated with magnetic resonance manipulation. 
However, there exists an essential  difference between the proposed system and other previously studied small quantum processors, such as NMR molecules. 
Here the boundary between the qubits in the system and the bath spins is ultimately dictated by our ability to effectively control the qubits. 
The interactions governing the couplings of the electronic spin to the nuclear qubit and bath spins are of the same nature, so that control schemes must  preserve the desired interactions among qubits while attempting to remove the couplings to the environment. The  challenges to overcome are then to resolve individual energy levels for qubit addressability and control, while  avoiding fast  dephasing due to uncontrolled portion of the bath. 

Before proceeding we note that various proposals for integrating these small quantum registers  into a larger system capable of fault tolerant quantum computation or communication have been explored  theoretically \cite{Jiang08,Campbell08} and experimentally \cite{BirnbaumEt,moehring07et}. These schemes  generally require a  \textit{communication} qubit and a few \textit{ancillary} qubits per register, in a hybrid architecture. The \textit{communication} qubits couple efficiently to external degrees of freedom (for initialization, measurement and entanglement distribution), leading to an easy control but also fast dephasing. The \textit{ancillary} qubits instead are more isolated and can act as memory and ancillas in error correction protocols. While our analysis is 
quite general in that it applies to a variety of  physical systems, such as quantum dots in carbon nanotubes \cite{Mason04} or spin impurities in silicon \cite{Morton08}, as a specific example we will focus on the nitrogen-vacancy (NV) centers in diamond \cite{Dutt06Et,Neumann08Et,HansonBathSciEt}. These are promising systems  for the realization of hybrid quantum networks due to their long spin coherence times and good optical transitions  that can be used for remote coupling between registers  \cite{jiang07et}. 
\begin{figure}[bht]
		\includegraphics[scale=.7]{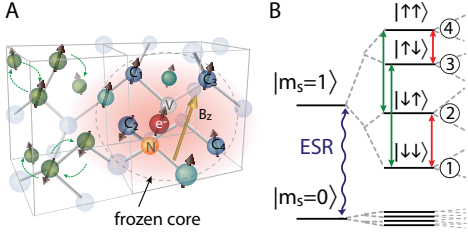}
	\label{fig:SpinModelNoisePaper}
	\caption{a) System model, showing the electronic spin in red and the surrounding nuclear spins. The closest nuclear spins are used as qubits. Of the bath spins, only the spins outside the frozen core  evolve due to dipolar interaction, causing decoherence. b) Frequency selective pulses, in a 3-qubit register.	}
\end{figure}

To be specific, we  focus on an electronic spin triplet, as it is the case for the NV centers. Quantum information is encoded in the $m_s$=$\{0$,$1\}$  Zeeman sublevels, separated by a large zero field splitting (making $m_s$ a good quantum number). Other Zeeman sublevels are shifted off-resonance by an external magnetic field  $B_z$, applied along the N-to-V axis.  The electron-nuclear spin Hamiltonian in the electronic rotating frame is 
\begin{equation}
	\label{spinHam}
	\begin{array}{l}
	\ham=\omega_L \sum I_z^j+S_z\sum A_j\cdot\vec{I}^j+\ham_\textrm{dip}\\
	=\frac{\openone-S_z}2\omega_L \sum I_z^j+\frac{\openone+S_z}2\sum \vec{\omega}_1^j\cdot\vec{I}^j+\ham_\textrm{dip}
	\end{array}
\end{equation}
where $S$ and $I^j$ are the electronic and nuclear spin operators, $\omega_L$ is the nuclear Larmor frequency, $A_j$ the hyperfine couplings and $\ham_\textrm{dip}$ the nuclear dipolar interaction, whose strenght can be enhanced by the hyperfine interaction \cite{Maze08b}. When the electronic spin is in the $m_s$=1 state, nearby nuclei are static (since distinct hyperfine couplings make nuclear flip-flops  energetically unfavorable in the so-called frozen core \cite{Khutsishvili69}) and give rise to a quasi-static  field acting on the electronic spin. The other bath nuclei cause decoherence via spectral diffusion \cite{witzelsousa,Maze08b}, but their couplings, which determine the noise strength and correlation time, are orders of magnitude lower than the interactions used to control the system. While in the $m_s$=0 manifold all the nuclear spins precess at the same frequency, the effective resonance frequencies in the $m_s$=1 manifold, $\omega_1^j$, are given by the hyperfine interaction and the enhanced g-tensor \cite{Childress06Et,Maze08b}, yielding a wide range of values. Some of the nuclear spins in the frozen core can thus be used as qubits. 
When fixing the boundary between system and environment we have to consider not only  frequency addressability  but also achievable gate times, which need to be shorter than the decoherence time. 

Control is obtained with  microwave ($\mu$w) and radio-frequency (rf) fields.  The most intuitive scheme, performing single-qubit gates with these fields and  two-qubit gates  by direct spin-spin couplings, is relatively slow, since rf transitions are weak. Another strategy, requiring only control of the electronic spin, has been proposed \cite{Hodges08,KhanejaESR}: switching  the electronic spin between its two Zeeman eigenstates induces nuclear spins rotations about two non parallel axes that generate any  single-qubit gate. This strategy is not  the most appropriate here, since rotations in the  $m_s$=0 manifold are slow \cite{endnote36}.
\begin{figure}[t]
\footnotesize
\[ 
\Qcircuit @C=0.5em @R=.3em {
\mathbf{A}\\
\lstick{\ket{e}\ \ } &\qw&\qw& & & & & \ctrl{1} &\gate{H}&\ctrl{1}&\gate{H}&\ctrl{2}&\gate{H}&\ctrl{1}&\gate{H}&\ctrl{1}&\qw\\
\lstick{\ket{C_1}}& \ctrl{1}&\qw& &\ \ =\ \ &  & & \gate{X}&\qw    &\gate{Z}&\qw     &\qw     &\qw     &\gate{Z}&\qw     &\gate{X}&\qw\\
\lstick{\ket{C_2}}& \gate{U}&\qw& & & & &\qw     &\qw    &\qw     &\qw     &\gate{U}&\qw     &\qw     &\qw     &\qw&\qw
}  
\]
\[
\Qcircuit @C=0.5em @R=.3em {
\mathbf{B}\\
\lstick{\ket{e}\ }& \gate{U}&\qw& & =& & &\gate{C}&\ctrl{1}&\gate{B}&\ctrl{1}&\gate{A}&\gate{X}&\ctrl{1}          &\gate{X}&\ctrl{1}&\gate{\alpha}&\qw\\
\lstick{\ket{C}}& \ctrl{-1}&\qw& & &  & &\qw     &\gate{Z}&  \qw   &\gate{Z}&\qw     &\qw        &\gate{\Phi_\alpha}&\qw     &\gate{\Phi_\alpha} &\qw&\qw
}
\]
\normalsize
\caption{Circuits for controlled gates $U$ among two nuclear spins (A) and a nuclear and the electronic spin (B). The gates $A, B, C$ are defined such that $U=e^{i\alpha}A Z B Z C$ and $ABC=\openone$, where $Z$ is a $\pi$ rotation around z \cite{IkeNielsen}. $\Phi_\alpha$ is a phase gate: $\ket{0}\bra{0}+\ket{1}\bra{1}e^{i\alpha/2}$ and the gate $\alpha$ indicates  $e^{i\alpha/2}\openone$.}
\label{circuits}
\end{figure}
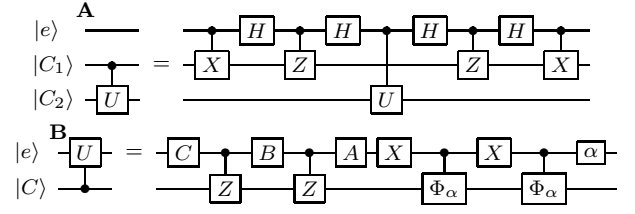
We thus propose another scheme to achieve universal control on the register, using only two types of gates: 1) One-qubit gates on the electronic spin and 
2) Controlled gates on each of the nuclear spins. 
The first gate can be simply obtained by a strong $\mu$w pulse. 
The controlled gates are implemented with rf-pulses on resonance with the effective frequency of individual nuclear spins in the $m_s$=1 manifold, which are resolved due to the hyperfine coupling and distinct from the bath frequency \cite{endnote37}. 
 Achievable rf power provides sufficiently fast rotations since the hyperfine interaction enhances the nuclear Rabi frequency when $m_s$=1 \cite{endnote38}. 
Any other gate needed for universal control can be obtained combining these two gates. For example, it is possible to implement a single nuclear qubit rotation  by repeating the controlled gate after applying a $\pi$-pulse to the electronic spin. 
Controlled gates between two nuclei  can  be implemented by taking advantage of the stronger coupling to the electron  as shown in Fig. \ref{circuits}(A). As long as the hyperfine coupling is several times larger than the nuclear coupling, the scheme avoiding any direct nuclear interaction is faster.
Although selectively addressing ESR transitions is a direct way to perform a controlled rotation with the electronic spin as a target, this is inefficient  as the number of nuclear spin increases.  The circuit in Fig. \ref{circuits}(B) performs the desired operation with only the two proposed gates on a faster time scale. 

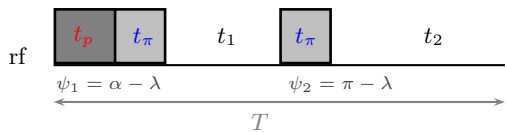
\begin{figure}[t]
\centering
\begin{pspicture}(-2, -.75)(7, .75)
  \psline{-}(0,-0.01)(6,-0.01)   \rput[bl](-.6,0){\small{rf}}
  \rput(2.25,.37){ $t_1$}  \rput(5,.37){ $t_2$}
  \psline[linecolor=gray]{<->}(0,-0.5)(6,-0.5)   \uput[ur](2.5,-1){{\gray $T$}}
  \psframe[linewidth=1pt,fillcolor=gray,fillstyle=solid](0,0)(.85,.75)  \rput(.4,.37){{\red $t_p$}}
  \psframe[linewidth=1pt,fillcolor=lightgray,fillstyle=solid](0.8,0)(1.5,.75)\rput(1.2,.37){{\blue $t_\pi$}}
  \psframe[linewidth=1pt,fillcolor=lightgray,fillstyle=solid](3,0)(3.7,.75)  \rput(3.36,.37){{\blue $t_\pi$}}
  \uput[ur](-.2,-.5){ {\scriptsize{\darkgray $\psi_1=\alpha-\lambda$}}}   
  \uput[ur](3,-.5){{\scriptsize{\darkgray$\psi_2=\pi-\lambda$}}}
\end{pspicture}
\caption{Rf pulse scheme for 1 nuclear spin gate  in the $m_s$=1 manifold. With  $t_p$=$\beta/\bar{\Omega}$ and fixing $\psi_1$=$\alpha$-$\lambda$ and $\psi_2$=$\pi$-$\lambda$, the time delays are  $t_1$=$\frac T2-t_p-t_\pi-\frac{\alpha+\gamma}{\omega}$ and $
t_2$=$\frac T2-t_\pi$+$\frac{\alpha+\gamma}{\omega}$.}
	\label{NVCnmrClock}
\end{figure}

When working in the $m_s$=1 manifold, each nuclear spin qubit is quantized along a different direction and we cannot define a common rotating frame. The evolution must be described in the \textit{laboratory} frame while 
the control Hamiltonian is  fixed in a given direction for all the nuclei (e.g. along the x-axis). 
This yields a reduced rf Rabi frequency  $\bar{\Omega}=\Omega_\textrm{rf}\sqrt{\cos{\varphi_1}^2\cos{\theta_1}^2+\sin{\varphi_1}^2}$ (where  $\{\theta_1,\varphi_1\}$ define  local quantization axes in the $m_s$=1 manifold and $\Omega_\textrm{rf}$ is the hyperfine-enhanced Rabi frequency). The propagator for a pulse time $t_p$ and  phase $\psi$ is
\[
	U_L(\Omega_\textrm{rf},t_p,\psi)=e^{-i [\omega t_p-(\lambda-\psi)]\sigma_{\tilde{z}}/2} e^{-i\bar{\Omega}t_p\sigma_{\tilde{x}}/2}e^{-i(\lambda-\psi)\sigma_{\tilde{z}}/2}
\]
where $\{\sigma_{\tilde{x}},\sigma_{\tilde{y}},\sigma_{\tilde{z}}\}$ are the  Pauli matrices in the local frame and  $\lambda$ is defined by $\tan{(\lambda)}=\tan{\varphi_1}/\cos{\theta_1}$. 
An arbitrary gate $U=R_{\tilde{z}}(\gamma)R_{\tilde{x}}(\beta)R_{\tilde{z}}(\alpha)$ can be obtained by combining $U_L$ with an echo scheme (Fig.~\ref{NVCnmrClock}), which  not only  refocuses the extra free evolution due to the lab frame transformation, but also sets the gate duration to any desired clock time common to all registers.  Fixing a clock time is advantageous to synchronize the operation of many registers.  This yields a minimum clock time  $T\geq4\pi\times\displaystyle{\textrm{Max}_{\bar{\Omega}_j,\omega_1^j}}\{\bar{\Omega}_j^{-1}+(\omega_1^j)^{-1}\}$.

In order to refocus the fast electronic-spin dephasing given by the frozen core nuclear spins, we need to embed the control strategy described above in a dynamical decoupling scheme \cite{DynDecViolaKnill} without loosing universal control, as explained in the following. 
 The electron-bath Hamiltonian is given by Eq. (\ref{spinHam}),  where the index $j$ now runs over  the spins in the bath. Neglecting for the moment the couplings among nuclei, we can solve for the evolution of the electronic spin under an echo sequence. 
By defining $U_0$ and $U_1$ the propagators in the 0 and 1 manifold respectively  
and assuming that the nuclear spins are initially in the identity state (high temperature regime), we  calculate the dynamics of the electronic spin, $\rho_e(t)=[\openone+(\ket{0}\bra{1}f_{ee}(t)+h.c.)]~\rho_e(0)$, where 
$f_{ee}(t)=\Tr{U_1U_0U_1^\dag U_0^\dag}$ $=\prod{[1-2\sin^2{(\theta_1^j)}\sin^2{(\omega_1^jt/2)}\sin^2{(\omega_Lt/2)}]}$ 
 is the function describing the echo envelope experiments \cite{ESEEM,Childress06Et}. 
Since in the $m_s$=$0$ subspace all the spins have the same frequency, $f_{ee}(2n\pi/\omega_L)=0$ and the electron comes back to the initial state.  
Nuclear spin-spin couplings lead to an imperfect echo revival and ultimately to decoherence via spectral diffusion \cite{witzelsousa,Maze08b}. The energy-conserving  flip-flops of remote nuclear spins modulate the hyperfine couplings, causing the effective field seen by the electron to vary in time. The field oscillations can be modeled by a classical stochastic process.
The overall evolution of the electronic spin is therefore due to two processes that can be treated separately as they commute:  the echo envelope calculated above and the decay due to a stochastic field, approximated by a cumulant expansion \cite{Kubo}.  
For a Lorentzian noise with autocorrelation function $G(\tau)=G_0 e^{-t/\tau_c}$ we obtain a spin-echo decay $\propto e^{-\frac{2\Omega^2}{3\tau_c}t^3}$ for $t\ll\tau_c$ (or  the motional narrowing regime for $\tau_c\ll t$). 

\begin{figure}[tb]
	\centering
	\psset{fillstyle=none}
\begin{pspicture}(-0.4, 0.3)(8, 2.2)
  \newrgbcolor{darkgreen}{0 100 0} 
  \psline{-}(0,.9)(7.6,.9)											\rput[bl](-.6,.9){\small{$\mu$w}}
  \psline{-}(0,1.7)(7.6,1.7)										\rput[bl](-.6,1.7){\small{rf}}
  \psline[linecolor=gray]{<->}(0,0.7)(7.6,0.7)	\rput(3.7,0.5){{\gray $T$}}
  \psframe[fillcolor=lightgray,fillstyle=solid](0,1.7)(.25,2.4)\rput(.09,1.5){{\footnotesize {\red t$_p$}}}
  \psframe[fillcolor=black,fillstyle=solid](.25,1.7)(.35,2.4)  \rput(.35,1.5){{\footnotesize{\blue $t_\pi$}}}
  \psframe[fillcolor=black,fillstyle=solid](1.9,1.7)(2,2.4)		 \rput(2,1.5)  {{\footnotesize{\blue $t_\pi$}}}
  \psframe[fillcolor=black,fillstyle=solid](3.7,1.7)(3.8,2.4)	 \rput(3.9,1.5){{\footnotesize{\blue $t_\pi$}}}
  \psframe[fillcolor=black,fillstyle=solid](5.7,1.7)(5.8,2.4)	 \rput(5.8,1.5){{\footnotesize{\blue $t_\pi$}}}
	\rput{0}(.67,2){{{\footnotesize $\tau$-t$_\varphi$}}}
	\rput(1.5,2)    {{\footnotesize $\tau$}}
	\rput(2.4,2)    {{\footnotesize $\tau$}}
	\rput(3.3,2)    {{\footnotesize $\tau$-t$_\varphi$}}
	\rput(4.3,2)    {{\footnotesize $\tau$+t$_\varphi$}}
	\rput(5.25,2)   {{\footnotesize $\tau$}}
	\rput(6.15,2)   {{\footnotesize $\tau$}}
	\rput{0}(7.2,2) {{\footnotesize $\tau$+t$_\varphi$-t$_p$}}
  \psframe(1.06,.9)(1.07,1.4) \psline[linestyle=dashed, dash=3pt 2pt,linewidth=.5pt]{-}(1.07,.9)(1.07,2.4)
  \psframe(2.87,.9)(2.88,1.4) \psline[linestyle=dashed,dash=3pt 2pt,linewidth=.5pt]{-}(2.87,.9)(2.87,2.4)
  \psframe(4.75,.9)(4.76,1.4) \psline[linestyle=dashed,dash=3pt 2pt,linewidth=.5pt]{-}(4.76,.9)(4.76,2.4)
  \psframe(6.65,.9)(6.66,1.4) \psline[linestyle=dashed,dash=3pt 2pt,linewidth=.5pt]{-}(6.65,.9)(6.65,2.4)
\end{pspicture}
\caption{rf and $\mu$w pulse sequence to implement a 1 nuclear spin gate while reducing the effects of a slowly varying electronic dephasing noise. The black bars indicate $\pi$-pulses, while the first rectangle indicate a general pulse around the x-axis. $\tau=\frac T8-\frac {t_\pi}2$ and  $t_\varphi=\frac{\alpha+\gamma}{4\omega}$.}
\label{PulseElectron}
\end{figure}
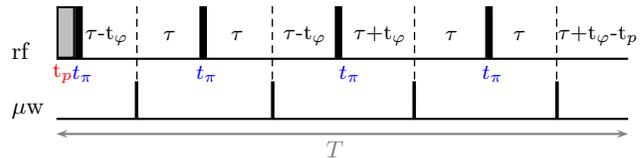

By using dynamical decoupling techniques \cite{LeakageCap} and selecting a cycle time multiple of the bare larmor precession period it is possible to extend the life-time of the electronic coherence. 
Figure \ref{PulseElectron} shows how to combine the electron modulation with the sequence implementing  spin gates. 
The effectiveness of these techniques relies on the ability to modulate the evolution on a time scale shorter than the noise correlation time. 
The noise of the electron-nuclear spin system is particularly adapt to these decoupling schemes. Consider for example the NV center. The measured electron dephasing $T_2^*$ time is about $1\mu s$ in natural diamonds \cite{Childress06Et}, as expected from MHz-strong hyperfine couplings. The intrinsic decoherence time $T_2$ can be orders of magnitude longer ($T_2\gtrsim 600\mu s$). This reflects the existence of a frozen region, where the spin flip-flops are quenched. The radius of this frozen core is about 2.2 nm \cite{Khutsishvili69} and only spins with hyperfine coupling $\lesssim$2.5 kHz contribute to  spectral diffusion. Both the inverse correlation time (given by the dipolar coupling among carbon spins) and the noise rms (given by the coupling to the electron) are of order of few kHz. Dynamical decoupling schemes must thus act on time scales of hundreds of $\mu s$. This in turns sets achievable constraints on the gate speed. 

The time of a conditional single nuclear spin rotation must be set so that the Rabi frequency $\Omega_\textrm{rf}$ is much less than the frequency splitting between two \textit{neighboring} spins (in terms of frequencies).  Suppose we want to control an $n$-spin register without exciting bath spins at the Larmor frequency. The minimum frequency splitting between two nuclei in the $m_s$=1 manifold will be at best $\delta\omega=\frac{\omega_M}{n}$ for nuclear frequency equally spaced and $\omega_M$  the maximum hyperfine-induced effective nuclear frequency.
The nuclear frequency spread due to the hyperfine interaction is then $\Delta E_N = \frac{n+1}2\omega_M$. We want the Rabi frequencies to satisfy the  constraints: 
$\Delta E_e\gg\Omega_e\gg\frac{n+1}2\omega_M>\frac{\omega_M}n\gg \Omega_\textrm{rf}$, 
where $\Delta E_e= 2g\mu_BB_z$ is the gap to other Zeeman levels ($m_s$=$-1$ for the NVc) and $\Omega_e$ the $\mu$w power. 
For a typical choice of 700G magnetic field along the NV axis, we have $\Delta E_e =
2$GHz and $\omega_L=0.8$MHz. Assuming $\omega_M \approx 20$MHz and $n = 4$ spins, we obtain the
following parameter window (in units of MHz)  
$2000 \gg \Omega_e \gg 25$, $\Omega_\textrm{rf} \ll 5$. 
 The gate clock time can be as short as a few $\mu s$, allowing hundreds of gates in the coherence time.

Since the scheme presented is based on selective pulses, the most important (coherent) errors will be due to off-resonance effects. If the Rabi frequency is much smaller than the off-set from the transmitter frequency $\delta\omega$, the off-resonance spin will just experience a shift (Bloch-Siegert shift) of its resonance frequency, $\Delta\omega_\textsc{bs}=\delta\omega-\frac 1t \int{\sqrt{\Omega_\textrm{rf}(t)^2+\delta\omega^2}~dt'}$ $\approx -\frac{\Omega_\textrm{rf}^2}{2\delta\omega}=-\frac{n\Omega_\textrm{rf}^2}{2\omega_M}$. This results in an additional phase acquired during the gate time that needs to be refocused.  Note that this error grows with the register size and constrains $\Omega_\textrm{rf}$. When reducing the Rabi frequency to achieve frequency selectivity, we have to pay closer attention to the rotating-wave approximation and consider its first order correction, which produces a shift of the on-resonance spin $\Delta\omega_{\textsc{rwa}}={\Omega_\textrm{rf}^2}/{4\omega_M}$. 
Other sources of errors are evolution of bystander nuclear spins and  couplings among  spins.
 More complex active decoupling schemes \cite{JonesKnill99,Leung00,Bowdrey05} can correct for these errors,  allowing to use the $m_s$=1 manifold as a  memory. 

Advanced techniques like shaped pulses, with  amplitude and phase ramping, composite pulses \cite{Levitt86}, pulses optimized via optimal control theory or with numerical techniques \cite{softpulsesEt,NavinGRAPEEt,Ryan08et} can provide better fidelity. The analytical model of control serves then as an initial guess for the numerical searches.
Pulses found in this way correct for the couplings among nuclei and  are robust over a wide range of parameters (such as experimental errors or the noise associated with static fields). Table  \ref{FidVsTimeGrape}  shows the results of simulations in a fictitious NV system with 1-4 nuclear spins and effective frequencies in the $m_s$=1 manifold ranging from 15 to 2MHz. 
We searched numerically via a conjugate gradient algorithm for a control sequence performing a desired unitary evolution, by varying the amplitude and phase of the $\mu$w and rf fields. 
We then simulated the control sequence  in the presence of noise, with contributions from a large, static field and a smaller fluctuating one. 
The projected fidelities in the absence of experimental errors are very high, a sign that the fast modulation of the electron evolution effectively decouples it from the bath. The pulse robustness  with respect to the noise is slightly degraded as the number of spins increases: the noise induces a spread of the electron resonance frequency, and it becomes more difficult to find a sequence of control parameters that drives the desired evolution in this larger Hilbert space for any of the possible electronic frequency. The fidelity degradation is however modest, and can be partially corrected by increasing the control field intensity. Furthermore, combining these pulses in a dynamical decoupling scheme would provide an efficient way to coherently control the registers. 
\renewcommand\arraystretch{1.2}
\begin{table}[htb]
	\centering
	\small
		\begin{tabular}{lcccc}
		 & {1 spin}& {2 spins}& {3 spins}& {4 spins}\\ %
		\hline  \\[-1.2em]\hline 	
		\multicolumn{1}{l}{$F$ (ideal) $\ \ $}	&$\ \ 0.9999\ \ $		& $\ \ 0.9999\ \ $& $\ \ 0.9992\ \ $				&$\ \ 0.9995\ \ $		\\
		\multicolumn{1}{l}{$F$ (noise)}	 	& 0.9994 	 			& 0.9995 				& 0.9975				&0.9925  		\\	
		\multicolumn{1}{l}{time }			 	& 5.0 $\mu$s 		  & 5.5 $\mu$s 		& 6.0 $\mu$s 	& 8.5 $\mu$s \\	  
		\hline
 		\end{tabular}
	\caption{Simulated fidelities ($F=|\Tr{U_w^\dag U}/2^N|^2$) at the optimal gate time in the presence of noise. The  gate is a $\pi/2$ rotation about the local x-axis for qubit 1 in a register of 1--4 nuclear qubits. The noise parameters are $T_2^*=1.5\mu$s and $T_2=250\mu$s; the maximum  Rabi frequencies are $2\pi\times10$MHz and $2\pi\times20$kHz for the electronic and nuclear spins respectively.  As expected, the optimal gate time increases with the register size, reflecting both the more complex control required in a larger Hilbert space and the weaker hyperfine couplings to more distant spins. Similar fidelities were obtained for a CNOT gate between spin 1 and 2.}
	\label{FidVsTimeGrape}
\end{table}

The size of the register is eventually limited by the number of available nuclear spins with a   hyperfine coupling strong enough to be separated from the bath. From experiments and ab-initio calculation \cite{Gali08} we expect hyperfine couplings of $\sim130$MHz in the first shell, and then a number of about 50 possible nuclear sites with hyperfine values from 15MHz to 1MHz. Even if the concentration of C-13 is increased (and  the Nitrogen nuclear spin is used) the size of the register will be limited to about 10 spins. Nevertheless such registers would be very useful for memory storage and error correction purposes.

In conclusion, we have presented a general approach to the control of a small quantum system comprising an electronic spin and few nuclear spins in the surrounding spin bath. We have shown that several of the bath spins can be isolated and effectively controlled, yielding a few-qubit register. These registers can be employed in many proposals for distributed quantum computation and communication, where coupling among registers could be provided either via photon entanglement \cite{jiang07et} or by a movable reading tip \cite{Maze08}. Our control methods enable algorithms  needed for error correction and entanglement purification, while the nuclear spins provide a long-time memory in the $m_s=1$ manifold, via  active refocusing, and the electron dephasing is kept under control by dynamical decoupling methods. We thus develop a modular control scheme, scalable to many registers and applicable to many physical implementations.

\textbf{Acknowledgments}.  
This work was supported by  the NSF, ITAMP, DARPA and the David and Lucile Packard Foundation.
\bibliographystyle{../../apsrev}

\begin{thebibliography}{36}
\bibitem{endnote36}{The rotation rates are faster for large external fields, which however reduce the angle between the two axes of rotations, thus increasing the number of switchings and the  gate time.}
\bibitem{endnote38}{The rf field, being perpendicular to the electronic zero-field splitting, is enhanced by second order electron-nuclear cross-transition \cite {AbragamBleaney}.}
\end{thebibliography}

\end{document}